\documentclass[12pt]{article}
\usepackage{amsmath}
\usepackage{bm}
\def\cA{{\cal A}}

\def\N{{I\kern-.25em{N}}}
\def\R{{I\kern-.25em{R}}}
\def\Z{{Z\kern-.5em{Z}}}
\def\V{{V\kern-.7em{V}}}

\def\d{\partial}
\def\bra{\langle}
\def\ket{\rangle}

\def\lr{{\rm L}^2({\rm R^d})}
\def\bvec#1{{\bm #1}}      
\def\vb{\bvec{b}}
\def\vk{\bvec{k}}
\def\vq{\bvec{q}}

\def\vx{\bvec{x}}
\def\dk#1#2{\frac{ d^{#2}{#1} }{ (2\pi)^{#2} }} 
\def\da#1#2{\frac{ d{#1}}{{#1}^{{#2}+1}}}
\begin{document}
\title{Langevin equation with scale-dependent noise} 
\author{M.V.Altaisky \\
Joint Institute for Nuclear Research, Dubna, Russia; and \\ 
Space Research Institute RAS, Profsoyuznaya 84/32, Moscow, 117997, Russia}
\date{March 21, 2003}
\maketitle
\begin{abstract}
A new wavelet based technique for the perturbative solution of 
the Langevin 
equation is proposed. It is shown that for the random force acting in a 
limited band of scales the proposed method directly leads to a finite 
result with no renormalization required. The one-loop contribution to 
the Kardar-Parisi-Zhang equation Green function for the interface growth is 
calculated as an example. 
\end{abstract}
The Langevin equation is one of the most general 
approximations for the evolution of a dynamical  system affected by 
fluctuating environment. It arises in the description of  magnetic at the 
presence of magnetic field fluctuations, in the description of 
hydrodynamic turbulence, in stochastic quantization problems,  
in the description of interface 
growth and in a large variety of other problems 
\cite{MSR1973,AVP1983,KPZ1986,ZJ1986}.

In the most general form the Langevin equation can be casted as
\begin{equation}
\frac{\d\phi(\vx,t)}{\d t} = U[\phi(\vx,t)] + \eta(\vx,t), 
\quad \bra \eta(x) \eta(x') \ket = D(x,x'),
\label{le1}
\end{equation}
where $U[\phi]$ is the nonlinear interaction potential, 
$\eta(t,\vx)$  is the 
Gaussian random noise, which standss for the fluctuations of the 
environment.The Minkovski-like  $(d\!+\!1)$-dimensional notation 
$x \equiv\! (\vx,t), k\equiv\!(\vk,\omega)$ is used hereafter. 

The standard    way to solve the Langevin equation \eqref{le1} 
is to introduce the small parameter $\lambda$ in the interaction potential $U$,
and then solve the system iteratively in each order of the perturbative 
expansion. 
The averaging over the Gaussian random force $\eta$ consists in evaluation 
of the pair correlators $\bra \eta \eta \ket$. 
The procedure is simplified by the assumption of the Gaussian statistics of 
the random noise which allows one to take into account only even order 
correlators of the random noise: all terms containing the odd 
number of $\eta$ are equal to zero. 
The diagram technique for the iterative solution of the Langevin equation is  
often called the Wyld diagram technique \cite{Wyld61}. 
Similarly to that in quantum field theory, it requires  
elimination of loop divergences by renormalization group methods 
\cite{AVP1983,KPZ1986}.

The structure of divergences arising in the perturbative solution of the 
Langevin equation depends on a particular type of the random force correlation 
function. Most approximations use $\delta$-correlated random force 
of the form 
\begin{equation}
\bra \eta(k_1) \eta(k_2) \ket = (2\pi)^{d+1} \delta(k_1+k_2) D(\vk_2). 
\label{cor1} 
\end{equation}
The spatial part of the correlation function $D(\vk)$ is assumed either 
to be a constant or to have a power behavior ($D(\vk)\sim |\vk|^{-\beta}$).

From the physical standpoint, a random force acting in a limited 
range of scales is often desirable.
For instance, a limited band forcing  is  
the case for the stirred hydrodynamic turbulence and stochastic interface 
growth. 
In this paper we propose a novel method to describe the limited band 
stochastic forcing. It is shown, that for a narrow band forcing, an 
appropriate choice of the  random force correlation functions yields a 
theory which is free of loop divergences. 
The proposed method preserves the whole structure 
of the perturbation expansion, with only the space of functions being changed.
In the limiting case of the scale independent forcing \eqref{cor1} all 
common results are preserved.

To study the dynamical system separately at each scale, following \cite{Alt1999}, 
instead 
of the usual space of the random functions  $f(x,\cdot)\in(\Omega, \cA, P)$, where
$f(x)\in \lr$ for each given realization of the random process, we turn   
to the multi-scale representation provided by the continuous wavelet transform 
\begin{equation}
W_\psi(a,\vb,\cdot) = \int |a|^{-\frac{d}{2}}
\overline{\psi\left(\frac{\vx-\vb}{a}\right)}f(\vx,\cdot)d^dx.
\label{wtr}
\end{equation}
The wavelet transform is 
performed here only in the 
{\em spatial} argument, but not in its 
temporal argument of $x$. This is because the structure of 
divergences and other important properties of the considered 
processes are determined by their spatial localization.

The existence an uniqueness of the inverse wavelet transform 
\begin{equation}
f(\vx,\cdot) = C_\psi^{-1} \int |a|^{-\frac{d}{2}} 
\psi\left( \frac{\vx-\vb}{a}\right) W_\psi(a,\vb,\cdot) \frac{dad\vb}{a^{d+1}}
\label{fa}
\end{equation}
is provided by the admissibility condition imposed on the basic wavelet
\begin{equation}
C_\psi = S_d^{-1} \int \frac{|\hat\psi(\vk)|^2}{|k|^d}d^dk 
= \int \frac{|\hat\psi(a\vk)|^2}{a}da 
< \infty.
\label{cpsi}       
\end{equation}
Factor $S_d$, the area of a unit sphere in $d$ dimensions, 
stands for the rotationally symmetric wavelet $\psi(\vx)=\psi(|\vx|)$.

The use of the wavelet images instead of 
the original stochastic processes provides an extra analytical 
opportunities. What is important for us, is the possibility 
of construction of more than one set of random functions 
$W(a,\vb,\cdot)$, the images of which  have 
coinciding correlation functions in the space of $f(x,\cdot)$.

It is easy to check, that the image of the correlation function of 
the processes  
$$
\bra \widehat W(a_1,k_1) \widehat W(a_2,k_2)\ket 
= C_\psi (2\pi)^{d} \delta^d(k_1+k_2)  
a_1^{d+1} \delta(a_1-a_2) D_0,   
$$
coincides with that of the white noise 
$$
\begin{array}{lcl}
\bra {\hat f}(k_1) \hat f(k_2)\ket &=& (2\pi)^{d} D_0 \delta^{d}(k_1+k_2) \label{gn}\\
\nonumber \bra \widehat W(a_1,k_1) \widehat W(a_2,k_2)\ket &=&  
 (2\pi)^{d} D_0 \delta^{d}(k_1+k_2)
(a_1 a_2)^{d/2} \overline{\hat \psi(a_1 k_1) \hat\psi(a_2 k_2)}.   
\end{array}
$$
Therefore, modeling the random force in the space 
of wavelet coefficients, we can provide a narrow band forcing  
keeping intact all required properties in the common space.

As an example, let us consider the well known Kardar-Parisi-Zhang model 
of interface growth \cite{KPZ1986}:
\begin{equation}
\dot Z -\nu \Delta Z = \frac{\lambda}{2} (\nabla Z)^2 + \eta .
\label{kpz}
\end{equation}
Substituting the wavelet transform in spatial argument   
\begin{equation}
Z(x)  = C_\psi^{-1} \int  \exp(\imath(\vk\vx-k_0 t)) a^{\frac{d}{2}}
\hat\psi(a\vk) \hat Z(a,k) \dk{k}{d+1} \da{a}{d}
\label{zt}
\end{equation}
into the equation \eqref{kpz}, with the random force of the form
\begin{equation}
\begin{array}{lcl}
\bra \widehat \eta(a_1,k_1) \widehat \eta(a_2,k_2)\ket 
&=&  C_\psi (2\pi)^{d+1} \delta^{d+1}(k_1+k_2)
a_1^{d+1} \delta(a_1-a_2)  D(a_2,\vk_2),\\ \bra \hat\eta(a,k) \ket &=& 0.
\end{array}
\label{sfnc}
\end{equation}
after straightforward calculations, we get an integral equation 
$$
\begin{array}{lcl}
(-\imath\omega + \nu\vk^2) \hat Z(a,k) &=& \hat\eta(a,k) - \frac{\lambda}{2} 
a^{\frac{d}{2}}  \overline{\hat\psi(a\vk)} 
C_\psi^{-2} \int (a_1 a_2)^\frac{d}{2}\hat\psi(a_1\vk_1)
\hat\psi(a_2(\vk-\vk_1)) \\ & & \vk_1  (\vk-\vk_1) 
\hat Z(a_1,k_1) \hat Z(a_2,k-k_1) \dk{k_1}{d+1} \da{a_1}{d}\da{a_2}{d},
\end{array}
$$
from where, the one loop approximation for the Green function is yielded 
\begin{equation}
G(k) = G_0(k) - \lambda^2 G_0^2(k) \int \dk{k_1}{d+1} \Delta(k_1)
\vk_1 (\vk-\vk_1)|G_0(k_1)|^2 \vk\vk_1 G_0(k-k_1) + O(\lambda^4)
\label{G2}
\end{equation}
where  $G_0^{-1}(k) = -\imath\omega + \nu\vk^2$ is the zeroth order 
approximation for the Green function, and 
\begin{equation}
\Delta(k) \equiv  C_\psi^{-1} \int \frac{da}{a} |\hat\psi(a\vk)|^2 D(a,\vk) 
\label{dak}
\end{equation} 
is the scale averaged effective force correlation function. 
The Green function obtained  with the random force \eqref{sfnc},
does not depend  on scale $\hat Z(a,k) = G(k) \hat \eta(a,k).$ 

Similarly, for the  pair correlation function $\bra Z Z \ket$, we get 
\begin{eqnarray}
 C(a_i,a_f,k) &=& \frac{\lambda^2}{2} |G_0(k)|^2 
\overline{\hat \psi(a_i\vk)\hat \psi(-a_f\vk)} \label{C2} \\ 
\nonumber & & \int \dk{k_1}{d+1} |G_0(k_1)|^2|G_0(k-k_1)|^2 
 [\vk_1(\vk-\vk_1)]^2 \Delta(k_1) \Delta(k-k_1). 
\end{eqnarray}
For the  random stirring which does not depend on scale, the integration 
in equation \eqref{dak}, after substitution $k_1=k_1'+\frac{k}{2}$ in both 
equations \eqref{G2}and \eqref{C2}, leads to the known result \cite{KPZ1986}. 
 
Let us consider a single band forcing 
\begin{equation}
D(a,\vk)=\delta(a-a_0) D(\vk)
\label{sb}
\end{equation}
and the ``Mexican hat'' taken as the basic wavelet 
\begin{equation}
\hat \psi(\vk) = (2\pi)^{d/2} (-\imath \vk)^2 \exp(-\vk^2/2), \quad 
C_\psi = (2\pi)^d.
\label{mh}
\end{equation} 
Substituting \eqref{sb} and \eqref{mh} into the equation \eqref{G2}, after 
integration 
over the frequency, in the leading order in the small parameter 
$x=|\vk|/|\vk_1|$,  we get the  contribution to the Green function ($d>2$):
\begin{equation}
G(k) = G_0(k) + \lambda^2 G_0^2(k) \frac{S_d}{(2\pi)^d}
\frac{a_0^3 k^2}{\nu^2}\frac{d-2}{8d} 
\int_0^\infty\!D(\vq) e^{-(a_0\vq)^2}  q^{d+1} dq + O(\lambda^4).
\label{gcor} 
\end{equation} 
For constant $D(\vq)\!=\!D_0$, the obtained contribution  
\label{g2mh} to the Green function is finite and does not require any further 
renormalization. In the limit $\omega,\vk\!\to\!0$, the one loop contribution 
to the surface tension $\nu$, which stems from \eqref{gcor}, is equal to  
\begin{equation}
\nu_{eff} = \nu\left[ 
1 - \frac{\lambda^2}{\nu^3k^2} \frac{d-2}{16d} \frac{S_d}{(2\pi)^d}
 a_0^{1-d} D_0 \Gamma\left(1+\frac{d}{2}\right)  + O(\lambda^4)  
\right].
\label{nueff}
\end{equation} 
We have to note that the equations (\ref{gcor},\ref{nueff}) were derived  
from one loop integral \eqref{G2} after substitution $k_1\!=\!q +k/2$, 
followed by perturbation expansion in small parameter $x=k/q$,
\cite{MHKZ1989}. For concrete correlation functions the direct 
numeric evaluation of the integral \eqref{G2} may be more adequate.
Similar calculations can be done for other interaction potentials.
For instance, for the quadratic interaction 
$\frac{\lambda^2}{2} Z^2$, the corresponding 
equations for \eqref{G2} and \eqref{C2} differ from that obtained above only 
by the 
absence of the scalar products of the wave vectors in each vertex. 

Concerning the contribution of the higher orders of the perturbation 
expansion, we should say that for the basic wavelets 
$\hat\psi(k)$ localized in $k$-space and limited band noise 
$D(a,k)$, the effective coupling constant $\bar\lambda$, which is 
the actual parameter of the perturbation expansion \cite{FNS1977}, 
can be made small by decreasing the noise amplitude. 
For instance, for the basic wavelets from the family 
$\hat \psi(k) = (2\pi)^{d/2} (-\imath k)^n \exp ( -k^2/2)$ and 
the noise correlator $D(a,k)\!=\!D_0\delta(a-a_0)$, the effective coupling 
constant is 
$$\bar \lambda^2 = \frac{\lambda^2}{\nu^3} \frac{D_0}{a_0}.$$

The author is grateful to N.Antonov and V.Priezzhev for 
useful discussions.


\begin{thebibliography}{9}
\bibitem{MSR1973} {\em Martin P.C., Sigia E.D., Rose H.A.} //
Phys. Rev. A. 1973. V.8. P.423--437.

\bibitem{AVP1983} {\em Adzhemyan L.Ts., Vasil'ev A.N., Pis'mak Yu. M. } //
Theor. and Math. Phys. 1983. V.57. P.1131-1143.

\bibitem{KPZ1986}{\em Kardar M., Parisi G., Zhang Y.-C.} //
Phys. Rev. Lett. 1986. V.56. P.889--892.

\bibitem{ZJ1986}{\em Zinn-Justin J.} //
Nucl. Phys. B. 1986. V.275. P.135--159.

\bibitem{Wyld61}{\em Wyld H.W.} //
Ann. Phys. 1961. V.14. P.143--165.

\bibitem{Alt1999}{\em Altaisky M.V.}// 
Eur. J. Phys. B. 1999. V.8. P.613--617.

\bibitem{MHKZ1989}{\em Medina E., Kardar, M., Parisi, G. and Zhang, Y.-C. } // 
Phys. Rev. A. 1989. V.39. P.3053--3075.

\bibitem{FNS1977}{\em Forster D., Nelson D.R., Stephen M.J.} // 
Phys. Rev. A. 1977. V.16. P.732--749.
\end{thebibliography}
\end{document}